# Fetus: the radar of maternal stress, a cohort study

Lobmaier SM*, Müller A*, Zelgert C, Shen C, Su PC, Schmidt G, Haller B, Berg G, Fabre B, Weyrich J, Wu HT, Frasch MG, Antonelli MC

**AFFILIATION:**

Lᴏʙᴍᴀɪᴇʀ SM (PD, Dr. med., MD, silvia.lobmaier@tum.de), Zᴇʟɢᴇʀᴛ C (cand. med, camilla.zelgert@web.de), Wᴇʏʀɪᴄʜ J (Dr. med., MD, joy.weyrich@yahoo.de): Department of Obstetrics and Gynecology, Klinikum rechts der Isar, Technical University of Munich, Germany

Mᴜ̈ʟʟᴇʀ A (MSc, alexander.mueller@mytum.de), Sᴄʜᴍɪᴅᴛ G (Prof. Dr. med., gschmidt@tum.de), Innere Medizin I, Department of Cardiology, Klinikum rechts der Isar, Technical University of Munich, Germany

Sʜᴇɴ C (BS, chao.shen@duke.edu), Sᴜ P-C (MD, b94401079@gmail.com), Department of Mathematics, Duke University, Durham, NC, 27705, USA

Hᴀʟʟᴇʀ B (Dr. rer. nat, bernhard.haller@tum.de), Institute of Medical Informatics, Statistics and Epidemiology

Bᴇʀɢ G (PhD, gaberg@ffyb.uba.ar), Universidad de Buenos Aires. Consejo Nacional de InvestigacionesCientíficas y Técnicas (CONICET). Facultad de Farmacia y Bioquímica. Buenos Aires, Argentina.

Fᴀʙʀᴇ B (PhD, brfabre2000@yahoo.com.ar), Universidad de Buenos Aires, Facultad de Farmacia y Bioquímica. Instituto de Fisiopatología y BioquímicaClínica (INFIBIOC). Buenos Aires.

Wᴜ H-T (MD, PhD, hauwu@math.duke.edu),Department of Mathematics and Department of Statistical Science, Duke University, Durham, NC, 27705, USA; Mathematics Division, National Center for Theoretical Sciences, Taipei, Taiwan.

Fʀᴀsᴄʜ MG (MD, PhD, mfrasch@uw.edu), Department of Obstetrics and Gynecology, University of Washington, Seattle, WA, USA.

Aɴᴛᴏɴᴇʟʟɪ MC (PhD, mca@fmed.uba.ar), Instituto de Biología Celular y Neurociencia "Prof. E. De Robertis", Facultad de Medicina, Universidad de Buenos Aires, Buenos Aires, Argentina.

\* equally contributed

Corresponding author:

PD Dr. Silvia M. Lobmaier

Ismaninger Str. 22

D- 81675 München (Germany)

Telefone number: +49 89 4140 5417

Fax: +49 89 4140 4892

Email: [silvia.lobmaier@tum.de](mailto:silvia.lobmaier@tum.de)

**Word count main text: 3349**

**Word count abstract: 250**

**SHORT TITLE**

Fetal heart rate reflecting prenatal stress

**ABSTRACT**

**Objective:** We hypothesized that prenatal stress (PS) exerts lasting impact on fetal heart rate (fHR). We sought to validate the presence of such PS signature in fHR by measuring coupling between maternal HR (mHR) and fHR.

**Study design:** Prospective observational cohort study in stressed group (SG) mothers with controls matched for gestational age during screening at third trimester using Cohen Perceived Stress Scale (PSS) questionnaire with PSS-10 ≥ 19 classified as SG. Women with PSS-10 < 19 served as control group (CG).

**Setting:** "Klinikum rechts der Isar" of the Technical University of Munich.

**Population:** Singleton 3rd trimester pregnant women.

**Methods:** Transabdominal fetal electrocardiograms (fECG) were recorded. We deployed a signal-processing algorithm termed bivariate phase-rectified signal averaging (BPRSA) to quantify coupling between mHR and fHR resulting in a fetal stress index (FSI). Maternal hair cortisol was measured at birth.

**Main Outcome Measures:** Differences for FSI between both groups.

**Results:** We screened 1500 women enrolling 538 of which 16.5 % showed a PSS-10 score ≥ 19 at 34+0 weeks. Fifty five women eventually comprised the SG and n=55 served as CG. Median PSS was 22.0 (IQR 21.0-24.0) in the SG and 9.0 (6.0-12.0) (p<0.001) in the CG, respectively. Maternal hair cortisol was higher in SG than CG at 86.6 (48.0-169.2) versus 53.0 (34.4-105.9) pg/mg (p=0.029). At 36+5 weeks, FSI was significantly higher in fetuses of stressed mothers when compared to controls [0.43 (0.18-0.85) versus 0.00 (-0.49-0.18), p<0.001].

**Conclusion:** Our findings show a persistent effect of PS affecting fetuses in the last trimester.

**Funding:** None

**Trial Registration:** NCT03389178.

**INTRODUCTION**

Prenatal exposure to maternal psychosocial stress, depression and anxiety confers lifelong risk for behavioral alterations that last beyond childhood [1]. Every fifth to fourth pregnant women experiences such prenatal stress (PS) [2]. Considering the difficulty to differentiate the concept of stress (including the psychological assessment of perceived stress) from anxiety, we will employ the term 'psychosocial stress' to refer to maternal general and pregnancy-specific stress and anxiety throughout this study [3,4]. The mechanisms of the physiological stress response can be divided into an acute response, which involves the rapid activation of the autonomic nervous system (ANS), and a delayed response mediated by the hypothalamic–pituitary–adrenal (HPA) axis. The neurally mediated ANS responses enable precise adjustments of target organs within seconds, while the HPA's slow response results in peak levels of cortisol within 10-30 min after acute stressor [5]. Most importantly, these maternal stress responses shape the development of the infant's stress response system, a phenomenon referred to as "fetal programming". It is the main mechanism by which prenatal exposures influence human postnatal development [6]. We hypothesized that the coordinated roles of the ANS and the HPA in the integrated stress response can be monitored non-invasively using electrocardiogram (ECG) and ECG-derived maternal and fetal heart rate (mHR, fHR). The relationship between mHR and fHR might provide important information about the functional status of fetal ANS [7]. In a prospective cohort of late-gestation women, we tested whether a new biomarker measuring the coupling of mHR and fHR can predict the preceding chronic exposure of mothers to stress. Herein, we propose a novel analysis method of coupling between mHR and fHR based on a signal-processing algorithm, first applied in adult cardiology, termed bivariate phase-rectified signal averaging (BPRSA) [8,9] and applied to trans-abdominally acquired fetal ECG (fECG). This method overcomes the limitation of non-stationary signal and background noise typical for fHR signal.

Such physiological biomarkers could serve as foundation for prediction of the child neurodevelopmental outcomes and aid in devising early developmental intervention strategies for children at risk of altered neurodevelopmental trajectories due to prenatal stress exposure. The aim of the study was to: 1) evaluate the coupling between mHR and fHR by BPRSA analysis; and 2) compare BPRSA results of fetal response to mHR changes in healthy controls and stressed fetuses.

**METHODS**

*Study design and study population:*

We performed a prospective observational cohort study in stressed mothers with controls matched for gestational age at screening. Subjects were recruited for 22 months (July 2016 until May 2018) from a cohort of pregnant women followed in the Department of Obstetrics and Gynecology at "Klinikum rechts der Isar" of the Technical University of Munich (TUM), a tertiary center of Perinatology located in Munich, Germany, serving ~2000 mothers/newborns per year.

*Experimental design:*

TUM obstetricians identified prospective subjects according to the following inclusion criteria: singleton pregnant women between 18 to 45 years of age in their third trimester (at least 28 weeks gestation). Exclusion criteria were a) serious placental alterations; b) fetal malformations; c) maternal severe illness during pregnancy and d) maternal drug or alcohol abuse.

The participants entered Phase I-III of the study (Figure S1):

**Phase I: Screening:**

We administered Cohen Perceived Stress Scale questionnaire (PSS-10). PSS-10 categorized them as SG for PSS-10 score ≥19 [2]. For every subject categorized as stressed, the next screened participant matching for gestational age at recording with a PSS-10 score < 19 was entered into Phase II as control.

**Phase II: Maternal and fetal prenatal ANS assessment:**

Prospective participants attended an informational session, when procedures were explained, formal enrollment completed and the consent forms from the participants obtained. We collected demographic information from the consented women.

Two and a half weeks after screening, we performed a transabdominal ECG (taECG) recording at 900 Hz sampling rate of at least 40 minutes duration using AN24 (GE HC/Monica Health Care, Nottingham, UK). We applied the fetal ECG extraction algorithm SAVER [10] to detect the fetal R-peaks and the maternal R-peaks in the taECG separately. With the fetal and maternal R-peaks, we obtained the fetal and maternal RR interval time series. The quality of taECG was estimated by the calculation of signal quality index (SQI) within windows of one second each. Regions where SQI was lower than 0.5 were marked as artifacts and were not considered for the analysis. Mean fetal heart rate and mean maternal heart rate were calculated. Mean maternal respiratory rate was derived from taECG and calculated according to Sinnecker et al [11].

**Phase III: Delivery:**

*Newborn recordings:* Clinical data including body metrics, pH and Apgar score, were recorded.

*Maternal cortisol assessment:* On the day of parturition, hair strands (~3 mm diameter) were collected from the posterior vertex region on the head as close to the scalp as possible [12]. Hair samples were sent to the Department of Clinical Biochemistry (Endocrinology Section) of the Faculty of Pharmacy and Biochemistry (University of Buenos Aires) for cortisol measurement using autoanalyzers. Based on an approximate hair growth rate of 1 cm per month, the proximal 3 cm long hair segment is assumed to reflect the integrated hormone secretion over the three-month-period prior to sampling. The 3 cm hair sample was wrapped in aluminum foil for protection and stored at room temperature up to three months. Fifty milligrams of hair obtained from the 3 centimeters closest to the roots (equivalent to 3 months of growth), were weighed in an analytical balance, as recommended by the Society of Hair Testing [12]. The weighted hair was cut with scissors into small pieces and put into a glass test tube with emery boards and 4 ml of methanol added. The tube was sealed with a cap and incubated in a shaking bath for 3 h at room temperature and then overnight at 50°C, for the steroid extraction. After incubation, the tubes were shaken 1 minute in a Vortex and 0.6 ml of the supernatant was collected in Khan glass tubes and evaporated until dryness at room temperature (24-48 h). The dry remnant was reconstituted with 300 μl in MD1 multidiluent for Immulite, provided by Siemens, and this sample was vortexed for 1 minute until its processing. The extracted cortisol was measured in Immulite 2000 ® automated CLIA analyzer as previously reported by our Laboratory for salivary cortisol [13]. This procedure has been validated with the standard method of mass spectrometry and was patented by the University of Buenos Aires [13].

*Measurement of maternal stress during pregnancy:*

Maternal psychosocial stress was measured using the Cohen Perceived Stress Scale (PSS 10). This questionnaire measures the degree to which situations in one's life are appraised as stressful and is a widely used psychological instrument to measure nonspecific perceived stress [14]. The PSS-10 predicts objective biological markers of stress and increased risk for disease among persons with higher perceived stress levels. Increased maternal prenatal stress, measured by PSS-10, was associated with temperamental variation of young infants and may represent a risk factor for psychopathology later in life [15]. The PSS-10 has been validated in German speaking populations and is a quick tool for screening stress among prospective subjects [16-18].

There is no recommended cut-point for high stress; and other studies have stratified this continuous variable using study-specific thresholds [2,19]. In our case, we performed a pilot study and we found that the 80% quantile of the PSS-10 was 19, which was then used as the cut-off score for stressed (SG) and control group (CG) in this study in accordance to prior studies [2,19].

**Bivariate phase-rectified signal averaging (BPRSA)**

The bivariate PRSA method is an extension of the "monovariate" PRSA method that we introduced for the analysis of fetal heart rate [20,21].

BPRSA allows for identifying and quantifying relationships between two synchronously recorded signals [8]. In this study these two signals are the maternal heart rate (mHR) as the trigger signal and the fetal heart rate (fHR) as the target signal.

The algorithm of BPRSA consist of four steps:

1.) At the beginning we identify all decreases in mHR and mark them as so called anchor points A.

2.) To investigate the response of the fetus at the defined anchor points the fHR is interpolated with a sample rate of 900 Hz as maternal ECG is registered with 900 Hz. Anchor points are identified by the time of occurrence within the fHR and are denoted as A'.

Segments of length *2L* around each anchor *A'* are selected in the fHR signal. In the current study we used L = 9000 samples, which correspond to a window of 20 seconds.

3.) All segments are aligned at the anchors leading to a phase-rectification of the segments. The BPRSA-signal *X* is obtained by averaging the aligned segments. Deflections in the BPRSA-signal can be interpreted as coupling between mHR and fHR.

4.) Finally, the BPRSA-signal *X* is quantified within a defined time span prior to and after the center of X. The time span starts 1.5 seconds after the center and ends at 2.5 seconds after the center. Therefore we use S1 and S2 as additional indices for the quantification. The data frame after the center is defined as L+S1 up to L+S2. On the other side, the data frame prior to the center is defined as L-S2 up to L-S1. Using S1 = 1350 and S2 = 2250 reflects the sample rate of 900 Hz in the BPRSA signal X and corresponds to 1.5 seconds and 2.5 seconds respectively. Coupling between mHR and fHR was analyzed by BPRSA resulting in a new parameter called fetal stress index (FSI). FSI is quantified by calculating the difference between the mean values of the data frames after and prior the center of X.

$$FSI = \frac{1}{S2-S1} \sum_{i=L+S1}^{L+S2} X(i) - \frac{1}{S2-S1} \sum_{i=L-S2}^{L-S1} X(i)$$

Note that the center of X (at indices L) corresponds to our anchor definition which was performed within the mRRI. Thus the FSI measures the response of the fetus on maternal heart rate decreases.

**Statistics**

Normal distribution was tested using Shapiro-Wilk test. For skewed distribution, medians and interquartile ranges were reported and for Gaussian distribution, mean and standard deviation. For categorical data, we show the absolute and relative frequencies. For comparison between groups, Mann-Whitney U tests, t-test for independent samples and Pearson Chi squared test were used. Receiver operating characteristics (ROC) analyses were performed to estimate the predictive performance of the quantitative variables for the presence of PS

For each fetus and mother, the fHR and mHR recorded at the same time were analyzed.

Finally, Pearson's correlation coefficient was used to evaluate the relationship between the fHR and mHR. All statistical tests were conducted two-sided and a p-value <0.05 was considered statistically significant for all comparisons. The fetal HR extraction algorithm was carried out in MatlabR2016a. Statistical analysis was performed using IBM SPSS Statistics for Windows, version 25 (IBM Corp., Armonk, N.Y., USA).

**Study approval and funding**

The study protocol is in strict accordance with the Committee of Ethical Principles for Medical Research from TUM and has the approval of the "Ethikkommission der Fakultät für Medizin der Technischen Universität München" (registration number 151/16S). ClinicalTrials.gov registration number is NCT03389178. Written informed consent was received from participants prior to inclusion in the study. There was no funding for the study. The project was developed and performed by own resources of Frauenklinik/Klinikum rechts der Isar.

## RESULTS

**Sociodemographic parameters and perinatal outcomes**

Between July 2016 and May 2018 we enrolled and followed 55 SG and 55 CG subjects. The cohort characteristics and perinatal outcome variables are summarized in Table 1.

Of all 1500 screened women, 538 returned the questionnaire and 16.5 % scored ≥ 19 on PSS-10 classifying as SG (Figure S2). Based on recruitment criteria, SG comprised 55 subjects and CG comprised 55 subjects at a similar median gestational age of 34.0 weeks. Median PSS of the SG was 22.0 (1st - 3rd quartile: 21.0-24.0) and that of the CG 9.0 (6.0-12.0) ($p<0.001$), respectively. The cortisol in maternal hair was 63% higher in SG versus CG ($p=0.029$) confirming the PSS results. The correlation of PSS and cortisol in hair was 0.182, no statistically significant association was observed ($p=0.098$). Area under the ROC curve for hair cortisol (prediction PS) was 0.639 ($p=0.029$).

The pre-gestational [24.2 (20.9-30.8) versus 21.5 (20.2-23.5), $p=0.001$] and at-study-inclusion [29.8 (26.0-36.7) versus 26.1 (24.5-28.7), $p=0.001$] median BMI of the SG patients was higher. Cord blood arterial pO2 was lower in the SG fetuses [17.9 (13.0-23.0) versus 21.1 (17.0-26.2) mmHg, $p=0.035$].

There were 9 times as many women in the SG diagnosed with gestational diabetes mellitus (GDM) than in the CG ($p=0.008$) and 5 times as many diagnosed with cases of autoimmune diseases ($p=0.014$).

Smoking was more frequent in the SG ($p=0.028$) and the number of planned pregnancies was lower in the SG than in the CG ($p=0.001$).

Less SG subjects visited university ($p=0.001$) and had the monthly net-income per household above 5000€ ($p=0.002$). More than two times as many SG than CG subjects ended up in cesarean delivery ($p=0.007$).

**BPRSA**

Data from 104 out of 110 subjects were used, because 6 subjects had poor ECG signal quality (2 CG fetuses and 4 SG).

Mean mHR, mean maternal respiratory rate calculated from mECG and median fHR were similar in both groups. Median FSI was significantly higher in SG compared to controls [0.43 (0.18-0.85) versus 0.00 (-0.49-0.18), p<0.001](Figure 1). That means that SG fetuses showed fHR decreases whereas CG fetuses remained "relaxed" after the maternal anchor "mHR decreases". This difference remained significant even after adjustment of relevant socioeconomic differences between both groups (BMI, university degree, household income> 5000€/month, smoking, planned pregnancy, diabetes, autoimmune diseases; p=0.012). Area under the receiver operating characteristics curve was 0.748 (p<0.001) (Figure 2). Figure 3 illustrates BPRSA results- the fetal response to maternal heart rate decreases.

Apart, we found a significant correlation of PSS and FSI (r=0.34, p<0.001).

# DISCUSSION

## MAIN FINDINGS

PS identified in the 3$^{rd}$ trimester impacts fetal physiology until delivery as evidenced by the altered fHR properties in fetuses of SG mothers and by the lower fetal oxygenation at birth.

The proposed BPRSA index (FSI) provides unique insights into the relationship between two biological systems - mother and fetus. In this study we could detect periodic mHR decreases reflecting typical pattern of maternal breathing (sinus bradycardia during expiration). Interestingly, CG fetuses remained "stable" during these periods whereas fetuses of stressed mothers showed significant decreases of fHR.

## INTERPRETATION

We hypothesize that this response is induced by the mechanical stimuli (diaphragm excursion which changes the uterine pressure). Future studies will need to address the question why CG fHR remained unchanged whereas SG showed decreased fHR during maternal breathing. A possible explanation might be the "over-sensitization" of SG fetuses' HPA axis or the differences in maturation of the sympathetic and parasympathetic branches of the ANS in contrast to the CG similar to data derived from animal models where lower weight twin sheep fetuses showed increased sympathetic activity and immaturity of circulatory control[22]. Physiologically, the ability of two complex weakly coupled systems to entrain each other is influenced by their intrinsic oscillatory properties and maturation, respectively. A well-studied example can be found in the cardiac pacemaker physiology [23].

Fetal ANS is very sensitive to maternal stress [24-26] and common markers of ANS such as fHR reactivity to a stimulus, reflect emerging individual differences in the development of the autonomic and central nervous systems related to styles of future emotional regulation and risk for psychopathology[27,28]. It is hence likely that the fHR response to mHR changes represents a fetal stress memory and may serve as a novel biomarker to detect PS effects early *in utero* which may help guide early interventions postnatally. For example, PS was linked to increased risk for autism spectrum disorder and alterations of ANS in autism spectrum disorders children have been reported [29,30].

We observed a mild fetal hypoxia at birth compatible with the concept of chronic reduction of uterine blood flow due to PS [31]. The result may be a reduced placental catecholamine clearance, thus elevating fetal catecholamine levels with hyperactive HPA axis and sympathetic nervous system. The postnatal developmental sequelae of these adaptations remain to be elucidated.

The persistent exposure to stress is validated on the maternal side by the higher maternal chronic cortisol levels at delivery in SG compared to CG. Perceived (subjective) measures of maternal stress did not correlate well with maternal cortisol at delivery in our study. This agrees with the body of inconclusive evidence seeking to link subjective stress exposure with cortisol values. Hair cortisol also appears to be a better biomarker than salivary cortisol for evaluation of the effectiveness of a stress reduction program in graduate students [32]. Meanwhile, other studies performed in women during or after pregnancy showed a lack of correlation. For example, in 768 mothers recruited shortly after delivery, hair cortisol did not correlate with self-reports of chronic stress, anxiety, or depressive symptomatology [33]. In a larger study involving 3039 pregnant women, maternal cortisol during pregnancy was mainly affected by biological and lifestyle factors, but not by psychosocial factors[34]. Moreover, hair cortisol and cortisone measured in pregnant women in their 2$^{nd}$ and 3$^{rd}$ trimester correlated with self-reported stress in both trimesters only with cortisol/cortisone metabolism, but not with cortisol alone [35]. While the lack of correlation may be due to questionnaire-based assessments or, the inherently high cortisol levels during pregnancy, the most likely cause is the presence of mechanisms mediating PS effects that do not alter maternal cortisol levels. fHR changes may be a good biomarker of such a mechanism as evidenced by our finding that FSI was more predictive of the subjective stress perception than cortisol. This result is in line with the psychological body of work such as the polyvagal theory [36] where heart rate variability has served as a good biomarker of internal emotional regulation which, in turn, reflects upon subjective coping with daily stress. Such complex relationships may represent physiologically more stable traits and hence correlate better than cortisol. Cortisol, even when measured from hair where it is supposed to serve as a stress memory of prior 2-3 months of exposure, remains still a single molecule biomarker. As such, it may be less capable of carrying a steady-state trait-like imprint than the FSI.

STRENGTHS AND LIMITATIONS

Our study has several strengths and limitations. One of the strengths is the prospective observational study design including women experiencing daily hassles stress rather than life events or natural disasters stress. We believe this makes the findings generalizable onto population of pregnant women seen in most antepartum follow-up centers. Our demographic data indicated distinct features of SG and CG regarding metabolic and socioeconomic status. This may be seen as a confounding factor for FSI, but even after adjustment for these possible confounders the difference of FSI between both groups remains significant. With an objective and low-cost biomarker of PS impact on mother and fetus in hand, prevention should be the subject of the future studies. A limitation of the study is that not all of the screened "control" women could be included into Phase II and III of the study due to limited "manpower". However, since all women qualifying as SG were enrolled, but not all controls, we suggest this had no impact on the reported differences due to PS.

OUTLOOK

Our findings warrant further investigations into the mechanisms of maternal-fetal stress transfer under the widely prevalent conditions of daily hassles (as opposed to extreme stress exposures), identification of epigenetic biomarkers impacting the stress axis and ANS activity and postnatal consequences of intrauterine imprinting of maternal stress upon fetal physiology. At present we are following up the offspring's neurodevelopment at 24 months and testing the feasibility of using a composite biomarker panel from salivary DNA as an early and non-invasive means in PS-exposed infants and fHR analysis to detect PS-induced ANS alterations that occur during fetal development [37].

**CONCLUSION**

In conclusion, we validated our hypothesis that PS-induced programming is reflected in mHR and fHR biomarkers of ANS activity. The biomarkers we identified can be harnessed for early detection and follow-up of children affected by PS. Early detection of altered neurodevelopmental trajectories opens new possibilities for designing more timely and effective interventions to improve outcomes of pregnancy affected by PS.

**Acknowledgment:**

MCA was awarded with the August Wilhelm Scheer Professorship Program (TUM) twice for a 6 months period stay at the Klinik und Poliklinik für Frauenheilkunde, Technische Universität München, Klinikum rechts-der Isar, Munich for the start-up of the FELICITy project.

**Disclosure of interests:**

HTW and MGF hold a provisional and a PCT patent on fetal ECG technology. All the other authors don´t report any financial, personal, political, intellectual or religious conflict of interest.

**Contribution to authorship:**

Lobmaier SM and Antonelli MC devised and proposed the study. Zelgert C, Weyrich J, Frasch MG, Wu HT contributed to design. Lobmaier SM, Müller A, Frasch MG, Wu HT, Shen, C, Su PC and Haller B undertook mathematical and statistical analysis and interpretation.

Lobmaier SM, Wu HT, Shen C, Su PC, Müller A, Schmidt G, Haller B, Berg G, Fabre B, Weyrich J, Zelgert C, Frasch MG, Antonelli MC contributed to data acquisition and drafted the submitted article and revised critically for important intellectual content.

**Ethics approval statment**

The study protocol is in strict accordance with the Committee of Ethical Principles for Medical Research from TUM and has the approval of the "Ethikkommission der Fakultät für Medizin der Technischen Universität München" (registration number 151/16S, June 15, 2016). ClinicalTrials.gov registration number is NCT03389178. Written informed consent was received from participants prior to inclusion in the study.

**Funding**

No funding

**Table/figure caption list:**

**Table 1:** Study outcome parameters

**Figure 1:** Box plots using bivariate phase-rectified signal averaging (BPRSA) method comparing controls (CG, blue) and stress group (SG, red)

**Figure 2:** Area under the ROC curve for prediction of maternal stress using bivariate phase-rectified signal averaging method

**Figure 3:** BPRSA analysis of fRRI. upper) PRSA signal X for maternal RR intervals. The anchor point definition, namely all heart rate deceleration, reflect the central oscillation of X. lower) The response of fetal RR intervals on the maternal decreases. The signal of the control BPRSA shows a significant lower response than the BPRSA curve for the fetus of a stressed mother. Also shown is the time span which is used for the quantification of FSI (yellow).

**Figure S1:** Study procedures

**Figure S2:** Recruitment chart

Table 1:

| Characteristics | Control | Prenatal stress | |
|---|---|---|---|
| | n=55 | n=55 | p |
| **Baseline** | | | |
| gestational age at screening (weeks) | 34.0 (33.4-335.0) | 34.0 (32.7-35.1) | 0.626 |
| gestational age at inclusion (weeks) | 36.7 (35.0-37.4) | 36.4 (35.4-37.4) | 0.844 |
| maternal age (years) | 35.2 (3.5) | 33.8 (5.4) | 0.108 |
| BMI pregestational (kg/m²) | 21.5 (20.2-23.5) | 24.2 (20.9-30.8) | 0.001* |
| BMI at inclusion (kg/m²) | 26.1 (24.5-28.7) | 29.8 (26.0-36.7) | <0.001* |
| european / caucasian | 50 (90.9) | 51 (92.7) | 0.728 |
| married | 41 (74.5) | 39 (70.9) | 0.669 |
| university degree | 45 (81.8) | 29 (52.7) | 0.001* |
| household income> 5000€/month | 35 (63.6) | 19 (34.5) | 0.002* |
| smoking | 1 (1.8) | 7 (12.7) | 0.028* |
| multiparity | 22 (40.0) | 30 (54.5) | 0.127 |
| planned pregnancy | 50 (91.0) | 36 (65.5) | 0.001* |
| IVF / ICSI | 6 (10.9) | 2 (3.6) | 0.142 |
| gestational diabetes | 1 (1.8) | 9 (16.4) | 0.008* |
| autoimmune disease | 2 (3.6) | 10 (18.2) | 0.014* |
| working status at screening | 2 (3.6) | 4 (7.2) | 0.388 |
| score PSS-10 | 9.0 (6.0-12.0) | 22.0 (21.0-24.0) | <0.001* |
| cortisol in maternal hair (pg/mg)[1] | 53.0 (34.4-105.9) | 86.6 (48.0-169.2) | 0.029* |
| maternal heart rate (bpm) | 87.0 (10.6) | 88.7 (9.3) | 0.382 |
| maternal respiratory rate | 27.9 (3.6) | 28.4 (3.6) | 0.437 |
| fetal heart rate (bpm) | 140 (136-146 | 140 (136-147) | 0.995 |
| FSI (ms) | 0.00 (-0.49-0.18) | 0.43 (0.18-0.85) | <0.001* |
| **Perinatal outcome** | | | |
| gestational age at delivery (weeks) | 40.0 (39.0-40.7) | 39.4 (38.6-40.6) | 0.058 |
| cesarean delivery | 10 (18.2) | 23 (41.8) | 0.007* |
| gender female | 24 (43.6) | 20 (36.45) | 0.436 |
| birthweight (g) | 3560 (412) | 3552 (470) | 0.922 |
| birthweight percentile | 52.1 (25.0) | 57.6 (25.7) | 0.270 |
| length (cm) | 53.0 (51.0-55.0) | 53.0 (52.0-55.0) | 0.591 |
| head circumference (cm) | 35.0 (34.0-36.0) | 35.0 (34.0-36.0) | 0.437 |
| Apgar min 5 | 10.0 (9.0-10.0) | 10.0 (9.0-10.0) | 0.173 |
| Apgar min 10 | 10.0 (10.0-10.0) | 10.0 (10.0-10.0) | 0.280 |
| 5-min Apgar<7 | 4 (7.2) | 2 (3.6) | 0.388 |
| admission to NICU | 2 (3.6) | 3 (5.4) | 0.647 |
| **Arterial cord blood analysis results** | | | |
| pH | 7.25(0.09) | 7.28 (0.08) | 0.137 |
| umbilical artery pH <=7.15 | 5 (9.1) | 3 (5.4) | 0.430 |
| base excess, mmol/l | -5.4 (3.4) | -4.9 (3.0) | 0.431 |
| pO2, mmHg | 21.1 (17.0-26.2) | 17.9 (13.0-23.0) | 0.035* |
| pCO2, mmHg | 51.2 (11.1) | 50.7 (9.4) | 0.826 |
| lactate mmol/l | 4.1 (1.4) | 4.0 (1.6) | 0.861 |
| glucose, mg/dl | 81.7 (20.8) | 75.5 (19.0) | 0.187 |

Data are median (interquartile range) or mean (SD) or n (%).

PSS: Perceived stress scale; BMI: Body-mass index; NICU: Neonatal intensive care unit;

* $p < 0.05$, [1]missing data of 14 CG and 17 SG; 0: control group (PSS<19) 1: stress group (PSS≥19)

Figure 1:

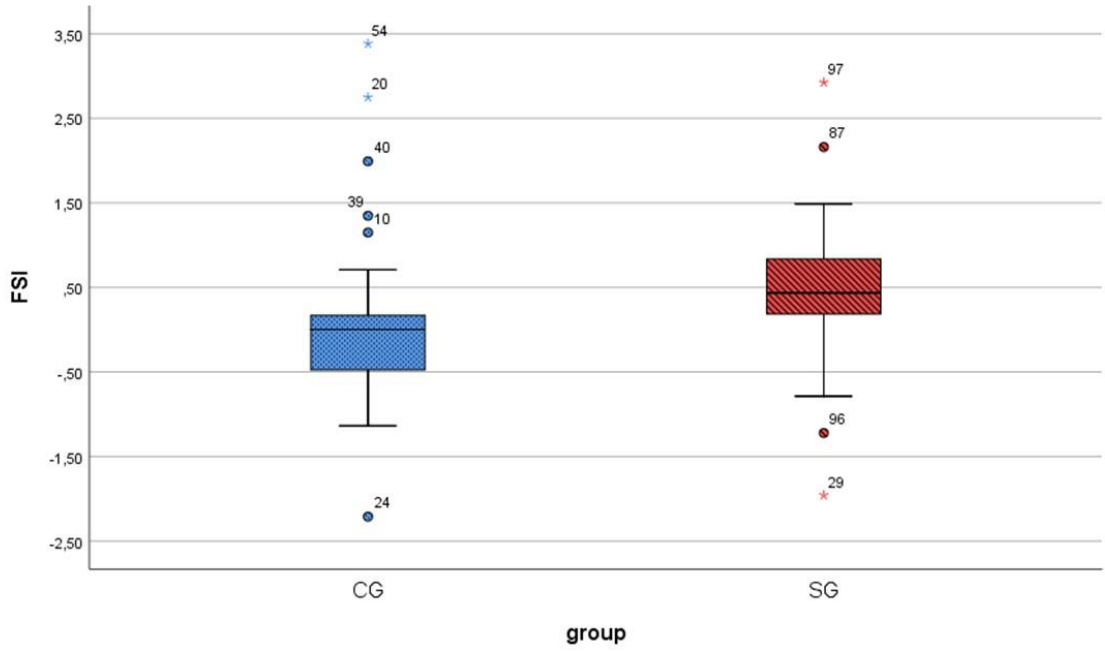

Figure 2:

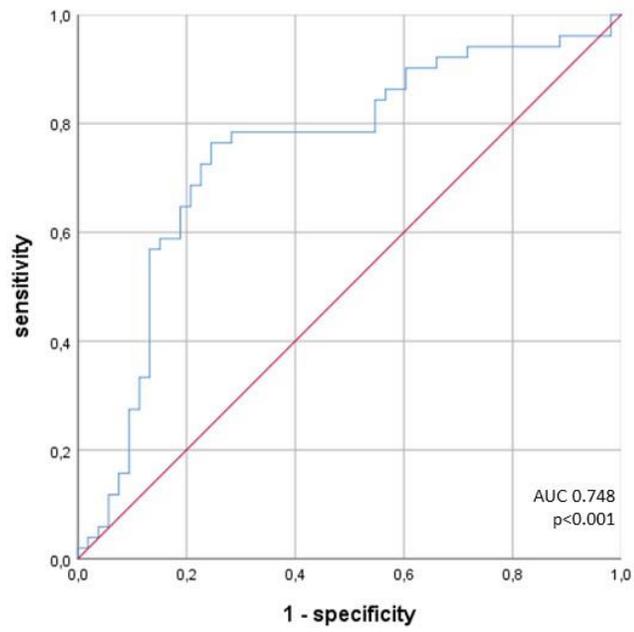

Figure 3:

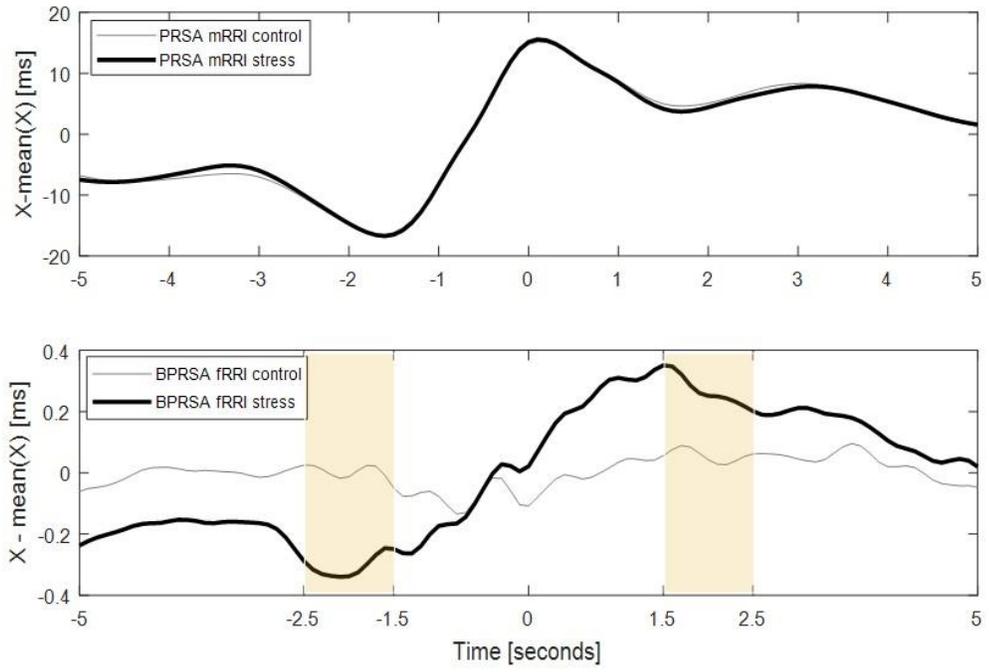

Figure S1:

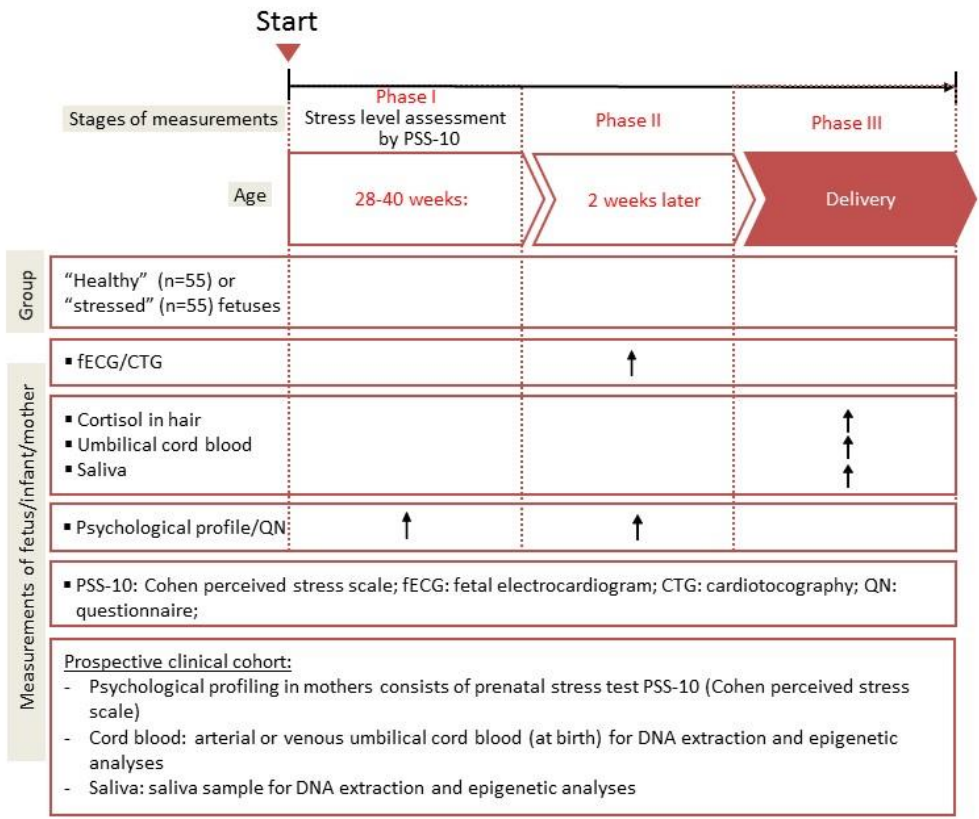

Figure S2:

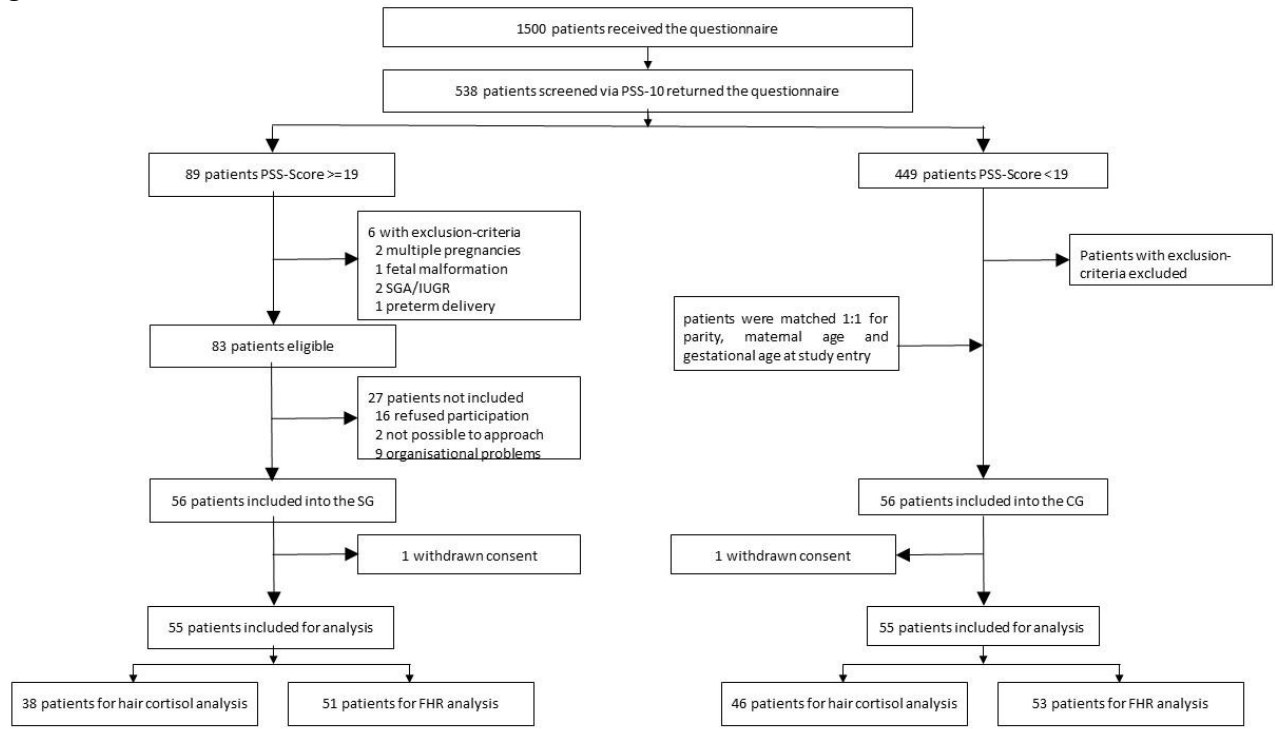

28